\documentclass{article}
\usepackage{amsmath}
\usepackage{amssymb}

\usepackage{graphicx}
\usepackage[normalem]{ulem}
\usepackage{overpic}
\usepackage[margin=1in]{geometry}

\begin{document}

\title{The Algorithmic Origins of Life}
\author{Sara Imari Walker$^{1, 2, 3}$ and  Paul C.W. Davies$^2$\\ \\
$^1$NASA Astrobiology Institute \\
$^2$ BEYOND: Center for Fundamental Concepts in Science \\
Arizona State University, Tempe, AZ\\ 
$^3$ Blue Marble Space Institute of Science, Seattle, WA \\
}
\date{}
\maketitle

\abstract{Although it has been notoriously difficult to pin down precisely what it is that makes life so distinctive and remarkable, there is general agreement that its informational aspect is one key property, perhaps the key property. The unique informational narrative of living systems suggests that life may be characterized by context-dependent causal influences, and in particular, that top-down (or downward) causation -- where higher-levels influence and constrain the dynamics of lower-levels in  organizational hierarchies -- may be a major contributor to the hierarchal structure of living systems \cite{Campbell, Campbell1990, Noble2012}. Here we propose that the origin of life may correspond to a physical transition associated with a shift in causal structure, where information gains direct, and context-dependent causal efficacy over the matter it is instantiated in. Such a transition may be akin to more traditional physical transitions ({\it e.g.} thermodynamic phase transitions), with the crucial distinction that determining which phase (non-life or life) a given system is in requires {\it dynamical} information and therefore can only be inferred by identifying causal architecture. We discuss some potential novel research directions based on this hypothesis, including potential measures of such a transition that may be amenable to laboratory study, and how the proposed mechanism corresponds to the onset of the unique mode of (algorithmic) information processing characteristic of living systems. }

\section{Introduction}

A landmark event in the history of science was the publication in 1859 by Charles Darwin of his book {\it On the Origin of Species} \cite{Darwin1859}, affording for the first time in history a scientific framework unifying all life on Earth under a common descriptive paradigm. However, while Darwin's theory gives a convincing explanation of how life has evolved incrementally over billions of years from simple microbes to the richness of the biosphere we observe today, Darwin pointedly left out an account of how life first emerged, ``One might as well speculate about the origin of matter'', he quipped. A century and a half later, scientists still remain largely in the dark about life's origins. It would not be an exaggeration to say that the origin of life is one of the greatest unanswered questions in science.
 
The origin of life constitutes three related but distinct questions: when, where and how did it happen? Progress toward understanding the first two has been markedly more successful than the third. Here we sidestep the when and where issues, which have been extensively discussed elsewhere (see {\it e.g.} \cite{Orgel1998a, Lazcano}), and address what is arguably the hardest and least constrained of the three origins questions: how did life begin? Of the many open questions surrounding how life emerges from non-life, perhaps the most challenging is the vast gulf between complex chemistry and the simplest biology: even the smallest mycoplasma is immeasurably more complex than any chemical reaction network we might engineer in the laboratory with current technology. The chemist George Whitesides, for example, has stated, ``How remarkable is life? The answer is: very. Those of us who deal in networks of chemical reactions know of nothing like it" \cite{Whitesides2004}. The heart of the issue is that we do not know whether the living state is ``just'' very complex chemistry, or if there is something fundamentally distinct about living matter. Right at the outset we therefore face a deep conceptual problem, one asked long ago by the physicist Erwin Schr\"odinger \cite{life}, namely, {\it What is Life}? Without a definition for life, the problem of how life began is not well posed.

Often the issue of defining life is sidestepped by assuming that if one can build a simple chemical system capable of Darwinian evolution then the rest will follow suit and the problem of life's origin will {\it de facto} be solved \cite{Joyce2012}. Although few are willing to accept a simple self-replicating molecule as living, the assumption is that after a sufficiently long period of Darwinian evolution this humble replicator will eventually be transformed into an entity complex enough that it is indisputably living \cite{Joyce2002a}. Darwinian evolution applies to everything from simple software programs, molecular replicators, and memes, to systems as complex as multicellular life and even potentially the human brain \cite{Fernando2012} -- therefore spanning a gamut of phenomena ranging from artificial systems, to simple chemistry, to highly complex biology. The power of the Darwinian paradigm is precisely its capacity to unify such diverse phenomena, particularly across the tree of life -- all that is required are the well-defined processes of replication with variation, and selection. However, this very generality is also the greatest weakness of the paradigm as applied to the origin of life: it provides no means for distinguishing complex from simple, let alone life from non-life. This may explain Darwin's own reluctance to speculate on the subject. 

Although it is notoriously hard to identify precisely what makes life so distinctive and remarkable \cite{Cleland2002, Tirard2010, Benner2010}, there is general agreement that its informational aspect is one key property, and perhaps the key property \cite{Szathmary1989, Kuppers1990, Yockey2005, HGCS}. The manner in which information flows through and between cells and sub-cellular structures is quite unlike anything else observed in nature. If life is more than just complex chemistry, its unique informational management properties may be the crucial indicator of this distinction, which raises the all-important question of how the informational properties characteristic of living systems arose in the first place.  This key question of origin may be satisfactorily answered only by first having a clear notion of what is meant by ``biological information''. Unfortunately,  the way that information operates in biology is not easily characterized \cite{Kuppers1990, JMS2000}. While standard information-theoretic measures, such as Shannon information \cite{Shannon}, have proved useful, biological information has an additional quality which may roughly be called ``functionality'' -- or ``contextuality'' -- that sets it apart from a collection of mere bits as characterized by Shannon Information content. Biological information shares some common ground with the philosophical notion of semantic information (which is more commonly -- and rigorously -- applied in the arena of ``high-level'' phenomena such as language, perception and cognition) \cite{JMS2000}. The challenge presented by requiring that one appeals to global context confounds any attempt to define biological information in terms of local variables alone, and suggests something fundamentally distinct about how living systems {\it process} information. In this paper, we postulate that it is the transition to context-dependent causation -- mediated by the onset of information control -- that is the key defining characteristic of life. We therefore identify the transition from non-life to life with a fundamental shift in the causal structure of the system, specifically, a transition to a state in which algorithmic information gains direct, context-dependent, causal efficacy over matter.

\section{Life and Information}

The concept of information has gained a prominent role in many areas of biology. Biologists routinely use terms such as ``signaling'', ``quorum sensing'', and ``reading'' or ``writing'' genetic information, while genes are described as being ``transcribed'', ``translated'', and ``edited'', all implying that in a biological context an informational narrative well captures the principal modes of activity and causal relationships. However, biological molecules are also physical objects, and like their non-biological counterparts, they may be described at the molecular (as opposed to system) level by a mechanical narrative. Obviously these parallel causal accounts, which appeal to different language and different concepts, must nevertheless fit together. Reconciling them is impeded by the fact that the precise nature of biological information remains vague and difficult to define (see {\it e.g.} \cite{SEP} for a detailed discussion). The information content of DNA, for example, is usually defined by the Shannon (sequential) measure. However, the genome is only a small part of the biological information story. DNA is not a blueprint for an organism\footnote{Whereas a blueprint provides a one-to-one correspondence between the symbolic representation and the actual object it describes, DNA does not contain all of the information necessary to reconstruct an organism \cite{Shea2007}. For example, many post-translational modifications as well as self-assembling components ({\it i.e.} lipids) are not encoded in the genome.}: no information is actively processed by DNA alone \cite{Noble2008}. Rather, DNA is a passive repository for transcription of stored data into RNA, some (but by no means all) of which goes on to be translated into proteins. The biologically relevant information stored in DNA therefore has very little to do with its specific chemical nature (beyond the fact that it is a digital linear polymer). The genetic material could just as easily be a different variety of nucleic acid (or a different molecule altogether), as recently experimentally confirmed \cite{XNA}. It is the functionality of the expressed RNAs and proteins that is biologically important. 

Functionality, however, is not a local property of a molecule \cite{Auletta2011}. It is defined only relationally, in a global context, which includes networks of relations among many sub-elements. For example, the functionality of expressed RNA and protein sequences is clearly context-dependent -- only an exceedingly small subset of these molecules are causally efficacious ({\it i.e.} meaningful) in the larger biochemical network of a cell whose functioning is dependent on conditions such as salinity of the cytoplasm, pH, {\it etc}. That milieu includes other expressed proteins, RNAs, metabolites, and a host of other molecules, the spatial distribution of which is crucial to their individual causal roles. {\it A priori}, it is not possible to determine which will be functional in a cell based on local structure and sequence information alone\footnote{While some algorithms are becoming efficient at predicting structure, biological functionality is always determined by insertion in a cell, or inferred by comparison to known structures.}. One is therefore left to conclude that the most important features of biological information ({\it i.e.} functionality) are decisively nonlocal. Biologically functional information is therefore not an additional quality, like electric charge, painted onto matter and passed on like a token. It is of course instantiated in biochemical structures, but one cannot point to any {\it specific} structure in isolation and say ``Aha! Biological information is here!''. Although the global and contextual nature of biological information has been widely recognized and discussed for some time \cite{Hogeweg2011}, we have only just begun unraveling the details of how cells (and larger organisms) organize and manage information \cite{Nurse2008}. The organization and information protocols for the epigenome and connectome, for example, remain little understood, and clearly involve a huge variety of regulatory RNAs and proteins. The recently published results of the ENCODE project, which provides an encyclopedia of human DNA elements \cite{ENCODE}, provides a glimpse of the complexity involved in mapping the {\it function} of the human genome.

Because complex non-linear systems are inherently prone to be unstable, organisms function only by being subject to regulation \cite{Doyle}, via a host of information control mechanisms that are themselves disseminated throughout the organism. It is the suite of regulatory molecules involved in information control that dictates the operating mode ({\it e.g.} phenotype) of a cell. Consider the genome and proteome systems. The current state -- {\it i.e.} the relative level of gene expression -- depends on the composition of the proteome, environmental factors, signaling molecules {\it etc.} that collectively act to up or down regulate individual genes. Such complex feedback loops serve to determine the future state of the system.  Linear causal chains are rarely apparent; rather causation is distributed throughout the state of the system as a whole (including information contained in the relations among all of the subcomponents).  Similar dynamics are at play throughout the informational hierarchies of biological organization, from the epigenome \cite{Davies2012}, to quorum sensing and inter-cellular signaling in biofilms, to the use of signaling and language to determine social group behavior \cite{Flack2007}.

In all of these cases where appeal is made to an informational narrative, we encounter context- (state-) dependent causation. In this respect, biological systems are quite unlike traditional mechanical systems evolving according to fixed laws of physics. In biological causation, subject to informational control and feedback, the dynamical rules will generally change with time in a manner that is both a function of the current state and the history of the organism \cite{Gould1979, Goldenfeld2011} (suggesting perhaps that even the concept of evolution itself may be in need of revision see {\it e.g.} Goldenfeld and Woese \cite{Goldenfeld2011, Goldenfeld2007} for an insightful discussion). A system in which the underlying rules (or laws) and the states both change with time in an interdependent way represents a decisive break with traditional  Newtonian dynamics, rooted in immutable laws, and opens up the possibility of novel pathways to emergent complexity that are as yet largely unexplored. 

The central position of information in biology is not itself especially new or radical \cite{Szathmary1989, Kuppers1990, Yockey2005}. What is often sidestepped, however, is the fact that in biological systems information is not merely a way to label states, but a property of the system. To be explicit, biological information is distinctive because it possesses a type of causal efficacy \cite{Auletta2008, Ellis2012} - it is the information that determines the current state and hence the dynamics (and therefore also the future state(s)).\footnote{The question of whether a causal chain expressed in informational language at the system level can ultimately be reduced, at least in principle, to a mechanistic causal chain at the molecular level, is the subject of a longstanding debate, complicated by the fact that biological systems are always open. We make no attempt to engage this notorious philosophical topic here, because it is irrelevant for the present discussion whether information is in fact a fundamental causal agent (which would represent a radical departure from standard physics), or may be treated merely phenomenologically as an effective causal agent.} We now turn to the question of how all this came about. How did information first gain causal purchase over certain complex systems that we now call living organisms? 

\section{Information in the Origin(s) of Life: Traditional Approaches}

A longstanding debate -- often dubbed the chicken or the egg problem -- is which came first, genetic heredity or metabolism \cite{Orgel1998a, Lazcano}? A conundrum arises because neither can operate without the other in contemporary life, where the duality is manifested via the genome/proteome systems. The origin of life community has therefore tended to split into two camps, loosely labeled as ``genetics--first'' and ``metabolism--first''. In informational language, genetics and metabolism may be unified under a common conceptual framework by regarding metabolism as a form of analog information processing (to be explained below), to be contrasted to the digital information of genetics. In approaching this debate, a common source of confusion stems from the fact that molecules play three distinct roles: structural, informational and chemical. To use computer language, in living systems chemistry corresponds to hardware and information ({\it e.g.} genetic and epigenetic) to software \cite{Davies1999}. The chicken-or-egg problem, as traditionally posed, thus amounts to a debate of whether analog or digital {\it hardware} came first.  

\subsection{A Digital Origin for Life}

The ``genetics-first'' paradigm, identifying a digital information repository as the most essential feature of the first living systems, is favored by biological approaches to the origin of life, which extrapolate backward in time from the properties of modern organisms. A widely accepted resolution to the seemingly inextricable duality of genotype/phenotype is that the modern ``DNA-protein'' world evolved from simpler precursor system involving only one major molecular species that played both the role of information carrier and of enzymatic catalyst. In modern organisms, RNA is a biochemical mediator, enabling the translation of DNA to protein. RNA is unique in that it can fill both roles, acting as both a genetic polymer and biochemical catalyst, with novel expanded roles for functional RNAs continually being discovered. This has led to the popular ``RNA world'' hypothesis, where all known life is posited to have descended from an ancestral population of organisms that utilized RNA as their sole major biopolymer prior to the advent of DNA and protein \cite{Gilbert1986, Cech1993, Joyce2002b, Joyce2010, RNAworld}. 

Despite the conceptual elegance of the RNA world, the hypothesis faces problems, primarily due to the immense challenge of synthesizing RNA nucleotides under plausible prebiotic conditions and the susceptibility of RNA oligomers to degradation via hydrolysis \cite{Levy1998, Shapiro2000, Sutherland2010}. Some of the chemical difficulties are alleviated if RNA was preceded by an alternative genetic polymer such as peptide nucleic acid (PNA) \cite{Nielsen} or threose nucleic acid (TNA) \cite{Orgel2000} (for other examples of candidate primitive genetic polymers see {\it e.g.} \cite{Eschenmoser2007}). In genetics-first origin of life scenarios, it has therefore been suggested that early-life may have undergone a ``hardware upgrade'' (or a succession of upgrades), eventually transitioning from a proto-RNA genetic polymer (or even an inorganic substrate \cite{CairnsSmith1982, Davies2004}) into a RNA-based biochemistry at a later stage in its evolutionary history. This system would then have undergone further hardware upgrades or ``genetic-takeovers'' to arrive at the DNA--protein world we observe today \cite{Leu2011}. 

However, beyond the chemical difficulties associated with synthesis and stability of primitive genetic polymers \cite{Engelhart2010}, lies a deeper conceptual challenge within the ``digital--first'' picture. As remarked above, the proteome -- and in fact nearly {\it all} biochemical interactions in the cell -- process information in an analog format, {\it i.e.} through chemical reactions which rely on continuous rates. For example, much of the information digitally stored in DNA must be first transcribed and translated before it becomes algorithmically meaningful in the context of the cell where it is then processed as analog information through protein interaction networks. Focusing strictly on digital storage therefore neglects this critical aspect of how biological information is {\it processed}. As we discuss below, due to the organizational structure of systems capable of processing algorithmic (instructional) information, it is not at all clear that a monomolecular system -- where a single polymer plays the role of catalyst and informational carrier -- is even logically consistent with the organization of information flow in living systems, because there is no possibility of separating information storage from information processing (that being such a distinctive feature of modern life). As such, digital--first systems (as currently posed) represent a rather trivial form of information processing that fails to capture the logical structure of life as we know it. 

\subsection{An Analog Origin for Life}

In contrast to models that rely on extrapolating backward in time from extant biology, approaches that move forward from what is known of the geochemical conditions on the primitive Earth typically favor an analog format for the first living systems. In analog chemical systems, information is contained in a continuously-variable composition of an assembly of molecules rather than in a discrete string of digital bits. ``Metabolism-first'' scenarios for the origin of life fall within this analog framework, positing that early life was based on autocatalytic metabolic cycles that would have been constructed in a manner akin to how analog computer systems are cabled together to execute a specific problem-solving task \cite{Dyson1982, Kauffman1993}.  The appeal of such metabolism-first scenarios is that the chemical building blocks - ranging from lipids \cite{Segre2001}, to peptides \cite{Huber98, Lee1996, Lynn2009}, to iron-sulfide complexes \cite{Wach, Russell1997} - are usually much easier to synthesize under abiotic conditions than any known candidate genetic polymer and would have therefore been much more abundant on the prebiotic Earth. The heritable information in this case typically consists of the compositional ratios of the molecules in the organized assemblies. Although it has been suggested that such ``composomes'' might provide a primitive inheritance mechanism \cite{Segre1999, Segre2000}, it is not clear that they are evolvable, since compositional information tends to degrade over successive generations inhibiting the capacity for open-ended evolution \cite{VSS2010} (see \cite{Markovitch2012} for a recent discussion of how such systems could be evolvable if possessing excess mutual catalysis). Therefore informational inheritance is not nearly as clear cut here as it is in the digital picture. 

Additionally, in the analog-first picture there exists a deeper issue of (re)programmability and with the difficulty of maintaining orthogonal ({\it i.e.} non-interacting and thus non-interfering) reactions in strictly analog reaction networks. Analog computers fell out of favor in the mid-20th century due to issues of universality - analog devices, regardless of their structure, are much more difficult to engineer to solve broad categories of problem than their digital counterparts. As we discuss below, all known life achieves universality (at least in a limited sense) by utilizing the {\it digital} sequence structure of informational polymers. Such universality would be exceedingly difficult to engineer in an analog-only system given the challenges associated with building reaction networks where each (programmed) reaction is chemically orthogonal to all other reactions. Orthogonality is, by comparison, relatively easy to achieve with digitized switches. Control is therefore much easier to achieve in an analog system with digital switches than in a solely analog system. Taking all of these factors into account, it is clear that analog-only systems are not capable of adaptation in the same way as living systems are. Modern life is a hybrid: digital memory and digital switches enable control over many (non-interfering) analog states, and therefore enable adaptability to changing environmental conditions with the same basic toolkit. This is another way of stating - in informational terms - that analog-only systems are not as versatile or as robust as analog systems with digital information control and as such may likely have very limited evolutionary capacity \cite{Doyle}.

\section{Redefining the Problem: An Algorithmic Origin for Life}

By the above considerations, it seems that digital or analog alone is insufficient to provide a satisfactory account of the origin of life -- not just on technical grounds, but for deep conceptual reasons. The former suffers from difficulties of prebiotic synthesis and due to fundamental limitations on how information can be processed in such scenarios (being trivial rather than nontrivial); whereas the latter suffers from issues of reprogrammability, control, and potentially long-term evolvability. This dilemma forms the crux of the chicken-or-egg problem cited above and suggests that focusing solely on the debate over chemical hardware may be limiting progress.  An implicit assumption of these traditional approaches has been that, while information may be manifested in particular chemical structures (digital or analog), it has no autonomy. As such, information -- though widely acknowledged as a key hallmark of life -- thus far, has played only a passive role in studies of life's emergence. Instead, hardware has dominated the discussion, in accordance with the generally reductionist flavor of biology in recent decades, with its associated assumption that, ultimately, all life is nothing but chemistry. 

However, as stressed above, a rigorous distinction between life and non-life is most likely to derive from the distinctive mode of information management and control displayed by living systems, {\it i.e.} that in biology information is causally efficacious. Both the traditional digital-first and analog-first viewpoints neglect the {\it active} (algorithmic or instructional) and distributed nature of biological information. In our view, an explanation of life's origin is fundamentally incomplete in the absence of an account of how the unique causal role played by information in living systems first emerged. In other words, we need to explain the origin of both the hardware and software aspects of life, or the job is only half finished. Explaining the chemical substrate of life and claiming it as a solution to life's origin is like pointing to silicon and copper as an explanation for the goings-on inside a computer. It is this transition where one should expect to see a chemical system literally take-on ``a life of its own'', characterized by informational dynamics which become decoupled from the dictates of local chemistry alone (while of course remaining fully consistent with those dictates). Thus the famed chicken-or-egg problem (a solely hardware issue) is not the true sticking point. Rather, the puzzle lies with something fundamentally different, a problem of causal organization having to do with the separation of informational and mechanical aspects into parallel causal narratives. The real challenge of life's origin is thus to explain how instructional information control systems emerge naturally and spontaneously from mere molecular dynamics. It is this issue which we explore in the remainder of this paper.

\section{Turing, von Neumann and Undecidability in the Origin of Life}

The instructional, or algorithmic, nature of biological information was long ago identified as a key property, and an early attempt to formalize it was made by von Neumann. He  approached the problem by asking whether it was possible to build a machine that could construct any physical system, including itself. Identifying the parallels between biological systems -- such as the human nervous system -- and computers, and drawing inspiration from Turing's work on universal computation, he sought a formalism that would include both natural and artificial systems \cite{VonN}. Turing showed that it was possible to build a device -- now known as a universal Turing machine -- which, given a sufficient amount of time, could output {\it any} computable function \cite{Turing}. A Turing machine is relatively simple hypothetical device, consisting of a machine and an unlimited memory capacity taking the form of an infinite tape marked out into squares, on each of which a symbol may be printed or erased, sequentially. A key feature of Turing machines is that both the state of the machine and the current symbol on the tape being read in, are necessary to determine the future evolution of the system. As such, the algorithm encoded on the tape plays a prominent role in the time evolution of the state of the machine. At least superficially, this appears to be very similar to the case presented by biological systems where the update rules change in response to information read out from the current state (as we discuss below, both are an example of top-down causation via information control). However, it is not obvious exactly how Turing's very abstract formalism might map onto biological systems. This was the problem von Neumann wished to solve. 

By analogy with Turing's universal machine, he therefore devised an abstraction called a universal-constructor (UC), a machine capable of taking materials from its host environment to build any possible physical structure (consistent with the available resources and the laws of physics) including itself. An important feature of UCs is that they operate on universality classes\footnote{Here we define a universality class as the set of all possible objects that can be made from a given set of building blocks}. In principle, a UC is capable of constructing any object within a given universality class (including itself, if it is a member of the relevant class). An example of such a universality class relevant to biological systems is the set of all possible sequences composed of the natural set of twenty amino acids found in proteins. The relevant UC in this case is the translation machinery of modern life, including the ribosome and associated tRNAs along with an array of protein assistants\footnote{The mapping between extant life and a von Neumann automaton is rather loose. In particular, the relevant UC here ({\it i.e.} the ribosome) is not included in the universality class it operates on and it therefore does not directly construct itself. There are a host of distributed control mechanisms and self-assembly processes that contribute to the reproduction of an entire cell.}. This system can, in principle, construct any possible peptide sequence composed of the coded amino acids (with minor variations across the tree of life as to what constitutes a coded amino acid \cite{Knight}). 

The UC forms the foundation of von Neumann's theory on self-replicating automata. However, a UC is a mindless robot, and must be told very specifically exactly what to do in order build the correct object(s). It must therefore be {\it programmed} to construct specific things, and if it is to replicate then it must also be provided with a blueprint of itself \footnote{Likewise a UC can construct {\it any} other object within its universality class if fed the appropriate instruction to do so.}. However, as von Neumann recognized, implicit in this seemingly innocuous statement is a deep conceptual difficulty concerning the well-known paradoxes of self-reference \cite{Hofstadter, Poundstone}. To avoid an infinite regress, in which the blueprint of a self-replicating UC contains the blueprint which contains the blueprint \ldots {\it ad infinitum}, von Neumann proposed that in the biological case the blueprint must play a dual role: it should contain instructions - {\it an algorithm} - to make a certain kind of machine ({\it e.g.} the UC) but should also be blindly copied as a mere physical structure, {\it without reference to the instructions its contains}, and thus reference itself only indirectly. This dual hardware/software role mirrors precisely that played by DNA, where genes act both passively as physical structures to be copied, and are actively read-out as a source of algorithmic instructions. To implement this dualistic role, von Neumann appended a ``supervisory unit'' to his automata whose task is to supervise which of these two roles the blueprint must play at a given time, thereby ensuring that the blueprint is treated both as an algorithm to be read--out and as a structure to be copied, depending on the context. In this manner, the organization of a von Neumann automaton ensures that instructions remain logically differentiated from their physical representation. To be functional over successive generations, a complete self-replicating automaton must therefore consist of three components: a UC, an (instructional) blueprint, and a supervisory unit. 

To rough approximation, all known life contains these three components, which is particularly remarkable, given that von Neumann formulated his ideas before the discoveries of modern molecular biology, including the structure of DNA and the ribosome. From the insights provided by molecular biology over the past 50 years, we can now identify that all known life functions in a manner akin to von Neumann automaton, where DNA provides a (partial) algorithm, ribosomes act as the core of the universal constructor and DNA polymerases (along with a suite of other molecular machinery) play the role of supervisory unit\footnote{An important note is that the all-important dual role cited above is clearly implemented: DNA polymerases are oblivious to the instructions that DNA contains and will blindly copy both coding and noncoding sequences.} \cite{Poundstone, Mange1998}. 

In spite of the striking similarities between a UC and modern life, there are some important differences. DNA does not contain a blueprint for building the entire cell, but instead contains only small parts of a much larger biological algorithm, that may be roughly described as the epigenetic components of an organism. The algorithm for building an organism is therefore not only stored in a linear digital sequence (tape), but also in the current state of the entire system ({\it e.g.} epigenetic factors such as the level of gene expression, post-translational modifications of proteins, methylation patterns, chromatin architecture, nucleosome distribution, cellular phenotype, and environmental context). The algorithm itself is therefore highly delocalized, distributed inextricably throughout the very physical system whose dynamics it encodes. Moreover, although the ribosome provides a rough approximation for a universal constructor (see footnote $^5$), universal construction in living cells requires a host of distributed mechanisms for reproducing an entire cell. Clearly in an organism the algorithm cannot be decomposed and stored in simple sequential digital form to be read out by an appropriate machine in the manner envisioned by Turing and von Neumann for their devices. 

Although the elements of von Neumann's UC cannot be put in a one-to-one correspondence with a living organism, the UC does provide a key insight into the nature of life, by directing attention to the logical structure of information processing and control, and information flow in living systems. 

\subsection{Trivial versus Nontrivial Self-Replication}

Although von Neumann automata are self-replicators, their mode of replication is non-trivial in a fundamental, logical, sense, and should be distinguished from trivial replicators such as crystals, viruses, computer viruses, nonenzymatic template replicators, lipid composomes, and Penrose blocks \cite{Penrose}.  Cast in the language of the previous section, trivial replicators process information strictly in the passive sense. Typically, they are characterized by building blocks which are not much simpler than the assembled object. Schr\"odinger recognized this key distinction in his take on {\it What is Life?} when he postulated that the genetic material must be some sort of ``aperiodic crystal'' \cite{life}. Algorithmic information theory can make the foregoing distinction precise. The algorithmic information of a system or structure is defined to be the Shannon information contained in the shortest algorithm that can specify the system or structure as its output \cite{Kolmogorov1965, Chaitin1969, Chaitin_AIT}. For example, a trivial replicator, such as a crystal, is one that may be specified by an algorithm containing far fewer bits than the system it describes. In contrast, a non-trivial replicator is algorithmically incompressible and requires an algorithm, or instruction set, of complexity comparable to the system it describes (or creates). 

A vast logical divide exists between trivial and nontrivial replicators because the former is not explicitly programmed. Instead, trivial replicators rely strictly on the implicit physics (and chemistry) of the current environment to support replication. Therefore only a limited set of objects within a given universality class is constructible. In other words, trivial self-replicating systems can only access one instructional mode -- the one which the system is currently operating in -- and as such are capable of only passive information handling. This stands in stark contrast to the case for nontrivial replicators, where {\it any} possible object within a given universality class (as defined above) -- {\it including the UC} -- can be constructed if the UC is provided with an appropriate instruction. Nontrivial replicators in some sense harness the underlying laws of physics and chemistry to achieve a broader agenda (although of course adhering to the constraints imposed by physical law). As such, only nontrivial replicators process information in an active sense, enabling the possibility for the update rules to change in response to the current informational state of the system (and vice versa). Because of this fundamental distinction in how information is handled and processed, nontrivial and trivial replication are two logically and organizationally distinct possibilities for self-replicating physical systems. The challenge in explaining life's origin is to account for the transition between trivial and non-trivial replication, which entails more than a mere leap in complexity, but a reconfiguration of the entire logical organization of the system.

\subsection{Algorithmic Takeover} 

Although modern life is clearly representative of the class of nontrivial self-replicators, the majority of work on the origin of life has focused on the conceptually simpler case of trivial self-replication. This not without good reason: the origin of translation -- mediating what is known of the transition from trivial to nontrivial\footnote{The informational narrative of life clearly goes beyond translation, however this is the one place in biology where we know universality (at least in a limited sense) has taken hold. A complete mapping of epigenetic factors will likely uncover other informational protocols at work in biological systems that may have some form of associated universality, and perhaps are even more primitive.} -- is notoriously difficult to pin down, amounting to an algorithmic takeover of information stored in one molecular species (nucleic acids) that becomes operable over another structurally and chemically very different species (peptides). The division of labor implicit in bimolecular life bestows one very obvious and distinctive advantage; it enables the instructions to be physically separated and stored away from the hardware that implements them. The ``arm's length'' control implicit in this division is exercised via a software channel - encoded transactions using messengers and specialized bilingual agents\footnote{``Bilingual'' here means tRNA molecules that recognize both the four-letter alphabet of nucleic acids and the twenty-letter alphabet of amino acids.} that identify, and are read by a system that can decode the instructions.  Thus the algorithm inhabits one molecular universe and its products inhabit another.  We consider this separation to be one of the hallmarks of life.

Although trivial self-replicators can undergo Darwinian evolution \cite{Eigen1971, Eigen1977a}, the lack of separation between algorithm and implementation implies that mono-molecular systems  are divided from known life by a logical and organizational chasm that cannot be crossed by mere complexification of passive hardware. In that respect we regard the case of the RNA world as currently understood as falling short of being truly living. If primitive ``life'' was strictly monomolecular, there would be no way to physically decouple information and control from the hardware it operates on, resulting in unreliable information protocols due to noisy information channels. For this rather deep reason, it may be that life had to be ``bimolecular'' from the start. 

We point out a curious philosophical implication of the algorithmic perspective: if the origin of life is identified with the transition from trivial to non-trivial information processing -- {\it e.g.} from something akin to a Turing machine capable of a single (or limited set of) computation(s) to a universal Turing machine capable of constructing any computable object (within a universality class) -- then a precise point of transition from non-life to life may actually be undecidable in the logical sense. This would likely have very important philosophical implications, particularly in our interpretation of life as a predictable outcome of physical law. 

\section{The Origin of Life: A Transition in Causal Structure}

We have argued that living and nonliving matter differ fundamentally in the way information is organized and flows through the system: biological systems are distinctive because information manipulates the matter it is instantiated in. This leads to a very different, context-dependent, causal narrative - with causal influences running both up and down the hierarchy of structure of biological systems ({\it i.e.} from state to dynamical rules and dynamical rules to the state) \cite{Campbell, Campbell1990, Noble2012, Davies2012}.  In modern life, genes may be up- or down-regulated by physical and chemical signals from the environment. For example, mechanical stresses on a cell may affect gene expression. Mechanotransduction, electrical transduction and chemical signal transduction -- all well-studied biological processes -- constitute examples of what philosophers term ``top-down causation'', where the system as a whole exerts causal control over a subsystem ({\it e.g.} a gene) via a set of time-dependent constraints \cite{Auletta2008, Levin2012, Davies2006}.  The onset of top-down information flow, perhaps in a manner akin to a phase transition, may serve as a more precise definition of life's origin than the ``separation of powers'' discussed in the previous section. The origin of life may thus be identified when {\it information gains top-down causal efficacy over the matter that instantiates it}. Top-down causation has an extensive literature so will not be reviewed here (see {\it i.e.} \cite{Campbell, Campbell1990, Noble2012, Auletta2008,  Ellis2012, Davies2006, Ellis2006, Ellis2011a}). 

 We note, however, that there may be several different mechanisms for top-down causation, which come into play at different hierarchical scales in nature \cite{Ellis2012}.  As we have presented it here, the key distinction between the origin of life and other ``emergent'' transitions is the onset of distributed information control, enabling context-dependent causation, where an abstract and non-physical systemic entity (algorithmic information) effectively becomes a causal agent capable of manipulating its material substrate \cite{Auletta2008, Ellis2012}. 

Although there is an extensive literature on top-down causation, particularly in biology, it has not been explicitly applied to the origin of life as such. The framework presented in this paper provides a well-defined definition for the transition to life, drawing on the top-down concept within an informational framework. Such a definition also addresses the vexed issue of what constitutes ``almost life''. This is essential for any theory that purports to chart a directional pathway from simple building blocks towards progressively more ``lifelike'' states. It makes sense to try to explain life's origin only if it resulted from processes of moderately highly probability, so that we can reasonably expect to explain it in terms of known science. It then follows from simple statistics that there will have been a large ensemble of systems proceeding down the pathway toward life, and no obvious reason why only one member successfully completed the journey. Ideally then, there should be a parameter, or more likely a set of parameters, to quantify progress toward life. The causal efficacy of distributed information control, discussed throughout this paper, provides a possible candidate parameter that includes the possibility of identifying states of ``almost life''. 
 
Walker {\it et al.} (WCD) have recently proposed via a toy model, one possible candidate measure for transitions in causal structure in biological hierarchies, using transfer entropy to study the flow of information from local to global and from global to local scales in a lattice of coupled logistic maps \cite{WCD}. Nontrivial collective behavior was observed to emerge each time the dominant direction of information flow shifted from bottom-up to top-down, indicating that top-down causation was in fact driving the emergence of collectives. The particular dynamical system investigated was designed to parallel a hallmark of many major evolutionary transitions -- the emergence of higher--level reproducers from previously autonomous lower-level units \cite{SzathmaryMS1995}. In this framework, the origin of life would mark the first appearance of this reversal in causal structure, and as such is a unique transition in the physical realm (marking the transition from trivial to nontrivial information processing as discussed earlier). The utility of this approach is that it provides a clear definition of what one should look for: a transition from bottom-up to top-down causation and information flow. 

The aforementioned simple model, while instructive, suffers from the fact that it cannot capture how algorithmic information alters the update rules, and thus the future state of the system.  A possible refinement is provided by Tononi's measure of so-called integrated information $\phi$, based on network topology \cite{Tononi}. This definition effectively captures the information generated by the causal interactions of the sub-elements of a system, beyond that which is generated independently by its parts. It therefore provides a measure of the distributed information generated by the network as a whole as a result of its causal architecture. Integrated information (also called ``excess information'') has recently been successfully applied to measure emergence in cellular automata under appropriate coarse-grainings of the dynamics \cite{Balduzzi2011}. A version of the theory whereby $\phi$ is in turn treated as a dynamical variable that then may influence the underlying causal relations among sub-elements might provide a way of quantifying the causal efficacy of information in the context discussed throughout this paper. 

\section{Conclusions}

\begin{table} \footnotesize
\centering
\begin{tabular}{ c  }
\hline
{\bf Hallmarks of Life} \\ \hline \hline
Global organization \\ 
Information as a causal agency \\ 
Top-down causation \\
Analog and digital information processing \\ 
Laws and states co-evolve \\ 
Logical structure of a universal constructor  \\ 
Dual hardware and software roles of genetic material \\ 
Non-trivial replication \\
Physical separation of instructions (algorithms) \\from the mechanism that implements them \\ 
\end{tabular}
\caption{The hallmarks of life.}
\label{tab}
\end{table}

We have presented a framework for understanding the origin of life as a transition in causal structure, and information management and control, whereby information gains causal efficacy over the matter it is instantiated in. The hallmarks of living systems based on this approach as discussed in this paper are summarized in Table \ref{tab}. The advantage of this perspective is that it provides a foundation for identifying the origin of life as a well-defined transition. In so doing, it forces new thinking in how life might have arisen on a lifeless planet, by shifting emphasis to the origins of information control, rather than -- for example -- the onset of Darwinian evolution or the appearance of autocatalytic sets ({\it i.e.} either analog or digital that lack information control), which, although certainly important to the story of life's emergence, do not rigorously define how/when life emerges as a function of chemical complexity. It also permits a broader view of life, where the same underlying principles would permit understanding of living systems instantiated in different chemical substrates (including potentially non-organic substrates). How this transition occurs remains an open question. While we have stressed that Darwinian evolution lacks a capacity to elucidate the physical mechanisms underlying the transition from non-life to life or to distinguish nonliving from living, evolution of some sort must still drive this transition (even if it does not define it). It is likely that nontrivial information processing systems with delocalized information are more evolutionarily robust given that information can be preserved in the face of changing environmental conditions due the physical separation of information and its material representation. 

Purely analog life-forms could have existed in the past but are not likely to survive over geological timescales without acquiring explicitly digitized informational protocols. Therefore life-forms that ``go digital'' may be the only systems that survive in the long-run and are thus the only remaining product of the processes that led to life. As such, the onset of Darwinian evolution in a chemical system was likely not {\it the} critical step in the emergence of life. As we have discussed, trivially self-replicating systems can accomplish this. Instead, the {\it emergence} of life was likely marked by a transition in {\it information processing} capabilities. This transition should be marked by a reversal in the causal flow of information from bottom-up only to a situation characterized by bi-directional causality. Methods to advance this program include identifying the causal architecture of known biochemical networks by applying candidate measures (such as $\phi$, or other measures of causal architecture \cite{Pearl2000, Crutchfield1994}), and focusing on regulatory networks (information control networks) in ancient biochemical pathways to identify the minimal network architectures necessary to support the causal and informational narrative observed in extant life.  A major unsolved problem is to determine how information control emerges {\it ab initio}, for example in an RNA world setting, from chemical kinetics, as well as how primitive control mechanisms might evolve and become increasing refined after ``algorithmic takeover'' has occurred.  Digitization may have been a natural outcome of this process in reaction-networks that had once been primarily analog. At this point, information would have become separated from its physical representation, permitting information to become a causal influence in its own right, and the language of Turing and von Neumann would have begun to apply. Characterizing the emergence of life as a shift in causal structure due to information gaining causal efficacy over matter marks the origin of life as a unique transition in the physical realm. It distinguishes nonliving dynamical systems, which display trivial information processing only, from living systems (and the complex systems derivative of biological systems, such as computers) which display nontrivial information processing as two logically and organizationally distinct kinds of dynamical systems.

\section*{Acknowledgments}
SIW gratefully acknowledges support from the NASA Astrobiology Institute through the NASA Postdoctoral Fellowship Program.  SIW also thanks the hospitality of the Aspen Center for Physics, supported in part by the National Science Foundation under Grant No PHY-1066293. PCWD was supported by NIH grant U54 CA143682.  We thank Andrew Briggs, Luis Cisneros, John Doyle and George Ellis for stimulating conversations as well as the manuscripts anonymous reviewers for constructive comments.

\bibliography{AlgorithmicRef}{}

\end{document}